\pdfoutput=1
\documentclass[5p]{elsarticle}
\usepackage{latexsym,amssymb,amsmath,mathrsfs,txfonts}
\usepackage{graphicx}
\usepackage{setspace,bm,color}
\usepackage[colorlinks=true, pdfstartview=FitV, linkcolor=red, citecolor=blue, urlcolor=blue]{hyperref}
\usepackage[caption=false]{subfig}

\newcommand{\diff}{\mathrm{d}}

\newcommand{\be}{\begin{equation}}      
\newcommand{\ee}{\end{equation}}      
\newcommand{\bea}{\begin{eqnarray}}      
\newcommand{\eea}{\end{eqnarray}}

\newcommand{\im}{\mathrm{i}}

\newcommand{\vphi}{\varphi}

\newcommand{\re}{{\rm Re}}
\newcommand{\iim}{{\rm Im}}

\begin{document}

\title{
{\vspace{-2.0cm}
\small\hfill\parbox{3.2cm}{\raggedleft%
RIKEN-QHP-202, RIKEN-STAMP-17
}}\\[1.5cm]
{\boldmath Complex saddle points and the sign problem in complex Langevin simulation}} 

\author[riken1]{Tomoya Hayata}
\author[riken2]{Yoshimasa Hidaka}
\author[riken2]{Yuya Tanizaki}

\address[riken1]{RIKEN Center for Emergent Matter Science (CEMS), Wako, Saitama 351-0198, Japan}
\address[riken2]{Theoretical Research Division, Nishina Center, RIKEN, Wako, Saitama 351-0198, Japan}

\date{\today}

\begin{abstract}
We show that complex Langevin simulation converges to a wrong result within the semiclassical analysis, 
by relating it to the Lefschetz-thimble path integral, 
when the path-integral weight has different phases among dominant complex saddle points.
Equilibrium solution of the complex Langevin equation forms local distributions around complex saddle points. 
Its ensemble average approximately becomes a direct sum of the average in each local distribution,
where relative phases among them are dropped.
We propose that by taking these phases into account through reweighting, we can solve the wrong convergence problem. 
However, this prescription may lead to a recurrence of the sign problem in the complex Langevin method for quantum many-body systems.

\end{abstract}

\maketitle

\section{Introduction}
Precise analysis of thermodynamic properties of a quantum many-body system, 
in particular, precise determination of its phase diagram is one of great challenges in theoretical physics.
An \textit{ab initio} simulation based on lattice field theory, in particular, so called Monte Carlo simulation is the most powerful tool for this.
In many interesting cases, however, Monte Carlo simulation is hindered by the notorious sign problem. 
The importance sampling, using the Boltzmann weight $\mathrm{e}^{-S}$, breaks down when the action becomes complex.
In hadron physics, lattice quantum chromodynamics (QCD) simulation suffers from the sign problem at finite quark densities~\cite{Muroya:2003qs,Aarts:2013lcm},
which is important to study quark matter inside neutron stars~\cite{RevModPhys.80.1455, Fukushima:2010bq}. 
The sign problem occurs also in condensed matter systems~\cite{RevModPhys.66.763,Sandvik,Pollet2012}.
Important examples are the fermionic Hubbard model away from half-filling, and geometric frustration in spin systems. 
A method to overcome the sign problem attracts a broad interest for application to the aforedescribed quantum many-body systems.

There have been a lot of attempts to tackle the sign problem. 
Among them, idea of complexification of the integration variables is one promising way to solve the sign problem.
Theoretical attempts along this line are classified into two approaches, that is, the Lefschetz-thimble and the complex Langevin methods. 
The Picard--Lefschetz theory gives a generalization of the steepest descent method, 
and Lefschetz thimbles are steepest descent paths in the extended complex plane~\cite{pham1983vanishing, Witten:2010cx, Witten:2010zr}. 
This method is formulated on rigorous mathematics, 
but it needs some approximation when applied to quantum many-body systems \cite{Cristoforetti:2012su, Mukherjee:2014hsa, Aarts:2013fpa, Fujii:2013sra,  DiRenzo:2015foa, Fukushima:2015qza}. 
On the other hand, the complex Langevin method is an extension of the Langevin equation to a complex Boltzmann weight~\cite{Parisi:1980ys,PhysRevA.29.2036,Parisi:1984cs,Damgaard:1987rr}.
The numerical implementation of this is possible based on lattice field theory. 
The complex Langevin method has been widely applied from condensed matter to hadron physics~\cite{Karsch:1985cb,Ambjorn:1986fz,Aarts:2008rr,Aarts:2008wh,Sexty:2013ica,Hayata:2014kra}.
There is a formal proof~\cite{Aarts:2009uq,Aarts:2011ax} on the correctness of the complex Langevin method, where it has been shown that the complex Langevin method correctly gives physical observables if the distribution obtained from the Langevin equation damps exponentially fast around infinities and singular points. 
This method is, however, known to give wrong results for some cases, where distribution does not show the exponentially fast decay, and thus the formal proof cannot be applied (For recent discussions, see also~\cite{Aarts:2010aq,Pawlowski:2013pje,Aarts:2013uza,Mollgaard:2013qra,Aarts:2014nxa,Nishimura:2015pba,Tsutsui:2015tua}).
Therefore, it is important to unveil what properties of the classical action cause the wrong convergence of the complex Langevin method.

In this paper, we show within the semiclassical analysis that complex Langevin simulation converges to a wrong result, 
when path-integral weight at complex saddle points has different phases.
This includes the case of the breakdown due to a singular drift term, e.g., the lattice QCD at finite density.
We reveal that complex Langevin simulation breaks down more generic case where the Langevin drift term has no singular point.
With the help of semiclassical analysis, we find that reweighting by the complex phase can partially solve the wrong convergence problem. 
However, the reweighting leads, in general, to a severe cancellation of the reweighting factor in many-body systems, 
which is nothing but a sign problem in terms of the complex Langevin method. 

\section{Complex Langevin method and its failure}\label{Sec:FailuerOfComplexLangevin}
For simplicity, we discuss an oscillatory integral of one variable $x$, which can be extended to multiple integrals in a straightforward way, 
\be
\langle O(x)\rangle={1\over Z}\int_{\mathbb{R}} \diff x\, \mathrm{e}^{-S(x)/\hbar} O(x). 
\label{eq:expectation_value}
\ee
where $Z$ is the normalization factor.
The action $S(x)$ is complex valued in general, which makes the Monte Carlo simulation of Eq.~(\ref{eq:expectation_value}) difficult because of the sign problem. 
One proposal to calculate Eq.~(\ref{eq:expectation_value}) for a complex valued action is the so-called complex Langevin method~\cite{PhysRevA.29.2036,Parisi:1984cs,Damgaard:1987rr}.
In this method, we solve the Langevin equations for complex values $z=x+iy$ along the fictitious time direction $\theta$,
\begin{eqnarray}
\partial_\theta z_{\eta}(\theta)= -S'(z_{\eta}(\theta)) +\sqrt{\hbar}\, \eta(\theta),
\label{eq:Langevin}
\end{eqnarray}
where $\eta(\theta)$ is real Gaussian noises satisfying  $\langle \eta(\theta)\rangle_\eta =0$, 
and $\langle \eta(\theta)\eta(\theta^\prime) \rangle_\eta =2\delta(\theta-\theta^\prime)$.
Since the action $S(x)$ is complex, the right-hand side of Eq.~\eqref{eq:Langevin} is also complex.
Thus, complexification of the variable $z$ is unavoidable, which is the reason why this method is called the complex Langevin method. 
In the real Langevin method i.e., if $S(x)$ is real, the ensemble average $\langle O(x_{\eta}(\theta))\rangle_\eta$ can be shown to converge to Eq.~(\ref{eq:expectation_value}) as $\theta\to \infty$. 
It has been shown that the complex Langevin method also converges to Eq.~(\ref{eq:expectation_value}) for the complex action $S(x)$ 
when the tail of distribution obtained from Eq.~\eqref{eq:Langevin} damps exponentially fast~\cite{Aarts:2009uq,Aarts:2011ax}. 
However, it has not been understood yet that what behavior is required to actions for the success of the complex Langevin simulation.
We first show based on the semiclassical analysis that the complex Langevin method gives wrong results if there are several dominant saddle points with different complex phases. 
After that, we propose a new prescription to evade this breakdown. 

Ito calculus shows the following: If the expectation value of a holomorphic operator $\widetilde{O}(z_{\eta}(\theta))$ converges as $\theta\to \infty$, 
the derivative of $\widetilde{O}(z_{\eta}(\theta))$, $O(z_{\eta}(\theta))=\widetilde{O}'(z_{\eta}(\theta))$, must satisfy the Dyson--Schwinger (DS) equation,
\be
\langle O(z_{\eta}) S'(z_{\eta})\rangle_{\eta}=\hbar\langle O'(z_{\eta})\rangle_{\eta}. 
\label{eq:holomorphic_DS_eq}
\ee
Here the argument $\theta= \infty$ of $z$ is omitted. 
When $\hbar=0$, the DS equation can be solved by complex saddle points $S'(z_{\sigma})=0$ ($\sigma \in \Sigma$). 
Even at finite $\hbar$, the contour integral on the steepest descent path $\mathcal{J}_{\sigma}$ around each $z_{\sigma}$ solves the DS equation~\cite{Pehlevan:2007eq,Guralnik:2009pk, Guralnik:2007rx},
\be
 \int_{\mathcal{J}_{\sigma}} \diff z \,\mathrm{e}^{-S(z)/\hbar} O(z)S'(z_{\eta})= \int_{\mathcal{J}_{\sigma}} \diff z \,\mathrm{e}^{-S(z)/\hbar} \hbar O'(z).
\ee
In general, any solutions of the DS equation are represented by a linear combination of  the contour integrals on the steepest descent paths~\cite{Guralnik:2007rx}. 
Therefore, ensemble average of a holomorphic operator $O(z_{\eta})$ at $\theta\rightarrow \infty$ can be represented as 
\be
\langle O(z_{\eta})\rangle_{\eta}={1\over Z}\sum_{\sigma\in\Sigma} d_{\sigma} \int_{\mathcal{J}_{\sigma}} \diff z \,\mathrm{e}^{-S(z)/\hbar} O(z). 
\label{eq:expectation_CL_Lefschetz}
\ee
Here, $d_{\sigma}$ is a complex number. 
The Lefschetz-thimble method~\cite{Witten:2010cx, Witten:2010zr, Tanizaki:2014xba} is useful to connect those steepest descent integrals with the original one~\eqref{eq:expectation_value}. 
If and only if $d_{\sigma}$ is an intersection number $\langle \mathcal{K}_{\sigma}, \mathbb{R}\rangle$ between a steepest ascent path $\mathcal{K}_{\sigma}$ and the original contour $\mathbb{R}$, the original integral (\ref{eq:expectation_value}) is recovered. 

We analyze Eq.~\eqref{eq:expectation_CL_Lefschetz} in the semiclassical limit $\hbar\to +0$. 
For this, we expand $S(z)$ around each complex saddle point $z_{\sigma}$ as  
\be
S(z_{\sigma}+\delta z)=S_{\sigma}+{\omega_{\sigma}\over 2}\delta z^2+O(\delta z^3). 
\ee
Let us first analyze the left hand side of Eq.~(\ref{eq:expectation_CL_Lefschetz}). 
If $\re\,\omega_{\sigma}>0$, the solution of the equation of motion (\ref{eq:Langevin}) can converge into $z_{\sigma}$ as $\theta\to \infty$ in $\hbar\to 0$. 
On the other hand, it cannot converge for $\re\,\omega_{\sigma}\le 0$. Then in the semiclassical approximation, we have
\be
\langle O(z_{\eta})\rangle_{\eta}\simeq \sum_{\sigma} c_{\sigma} O(z_{\sigma}), 
\label{eq:expectation_CL_Lefschetz_LHS}
\ee
where $c_{\sigma}\ge 0$, and $c_{\sigma}=0$ if $\re\,\omega_\sigma\le 0$. 
Next let us analyze the right hand side of Eq.~(\ref{eq:expectation_CL_Lefschetz}). 
In the semiclassical approximation, the integral along the thimble becomes 
\be
\int_{\mathcal{J}_{\sigma}} \diff z\; \mathrm{e}^{-S(z)} \mathcal{O}(z)
\simeq
\sqrt{2\pi \hbar\over \omega_{\sigma}}\mathrm{e}^{-S_{\sigma}/\hbar}\mathcal{O}(z_{\sigma}). 
\label{eq:expectation_CL_Lefschetz_RHS}
\ee
Also the denominator $Z\simeq Z_{\rm semi}$ can be evaluated using the semiclassical analysis 
by setting $\mathcal{O}(z_{\sigma})=1$ in the above discussions.
Now, from the comparison of the both sides of Eq.~(\ref{eq:expectation_CL_Lefschetz}) for arbitrary operators, we reach
\be
c_{\sigma}={1\over Z_{\rm semi}}\sqrt{2\pi \hbar\over \omega_{\sigma}}\mathrm{e}^{-S_{\sigma}/\hbar} d_{\sigma}
\label{eq:expectation_CL_Lefschetz_LHS_coefficient}
\ee
for dominantly contributing saddle points.
Note that $c_{\sigma}\ge 0$. However, $d_{\sigma}$ needs to be an integer $\langle \mathcal{K}_{\sigma},\mathbb{R}\rangle$ to recover \eqref{eq:expectation_value}. 
These two statements contradict with one another in general. 
As a result, we can conclude the following at least for semiclassical analysis: 
The complex Langevin method cannot reproduce the original integral~\eqref{eq:expectation_value}
if there are several dominant saddle points with different complex phases. 

Equation (\ref{eq:expectation_CL_Lefschetz_LHS_coefficient}) is completely sure for dominant saddle points. 
The above contradiction must be taken into account if the dominant saddle points have different complex phases. 
In other words, complex Langevin method may fail if there is some relative phase between the dominant saddle points.
For subdominant saddle points, the ambiguity of Borel resummation of large order perturbations can give nontrivial cancellations~\cite{Bogomolny1980431, Zinn-Justin1981125, Basar:2013eka, Cherman:2014ofa, Dorigoni:2014hea, Behtash:2015kna, Behtash:2015kva, Behtash:2015loa}. 
Therefore, we cannot judge from our argument whether the complex Langevin method gives a correct result 
if there is only one dominant saddle point. 
This subtlety needs further studies. 
For a Gaussian action, it is easy to check that Eq.~(\ref{eq:expectation_CL_Lefschetz_LHS_coefficient}) is satisfied and the complex Langevin method works well. On the other hand, there exists a model with power-law tail, 
where the complex Langevin method does not work but there is only one dominant saddle point (see, e.g., \cite{Aarts:2014nxa}). 

Let us give a few comments on previous studies. There is a formal proof~\cite{Aarts:2009uq,Aarts:2011ax} on the correctness of the complex Langevin method, but it relies on several nontrivial assumptions\footnote{One of the most nontrivial assumptions would be the $C_0$ semigroup property generated by Fokker--Planck-type partial differential operators. }. 
Combined with a recent study \cite{Nishimura:2015pba}, they have shown that the formal proof breaks down if the complex Langevin distribution does not decay exponentially fast around infinities and singular points.
Our analysis suggests without accessing details of the complex Langevin distribution that the breakdown happens if the dominant complex saddle points have different phases.
Even in many-body systems, we can obtain saddle points by numerically solving Eq.~\eqref{eq:Langevin} without random noises.
This situation would naturally bring us to the conjecture that the complex Langevin distribution has a polynomial tail around infinities or singular points if several dominant saddle points contribute with different phases. 
It would be an important future study to check this conjecture in order to achieve a deeper understanding of the complex Langevin method. 

\section{Prescription}
Let us propose a prescription to circumvent this inconsistency. 
We denote the equilibrium distribution of the complex Langevin method by $P$. The expectation value is given as 
\be
\langle O(z_{\eta})\rangle_{\eta}=\int\diff x\diff y \,P(x,y)O(x+\im y).
\ee 
In the semiclassical limit, $P$ will be represented, by using a sum of distributions $P_{\sigma}$ localized at complex saddle points $z_{\sigma}$, as 
$P=\sum_{\sigma} P_{\sigma},$
which gives the expectation value (\ref{eq:expectation_CL_Lefschetz_LHS}) i.e, 
\be
\int\diff x\diff y\, P_{\sigma}(x,y) O(x+\im y)\simeq c_{\sigma} O(z_{\sigma}).
\ee 
By defining (nonholomorphic) functions $\chi_{\sigma}$ satisfying $\chi_{\sigma}P_{\tau}\simeq \delta_{\sigma \tau}P_{\tau}$, we define a phase function $\Theta$ by 
\be
\Theta(z,\overline{z}):= {1\over Z}\sum_{\sigma}{1\over c_{\sigma}}\langle \mathcal{K}_{\sigma},\mathbb{R}\rangle\sqrt{2\pi \hbar\over \omega_{\sigma}}\mathrm{e}^{-S_{\sigma}/\hbar} \chi_{\sigma}(z,\overline{z}).
\ee 
If $P_{\sigma}$ does not overlap with others, $\chi_{\sigma}$ can be chosen as the characteristic function of $\mathrm{supp}(P_{\sigma})$. 
This is not true in general, and we must find $\chi_{\sigma}$ satisfying the condition with a good approximation. 
The expectation value of a holomorphic operator $O(z)$ is given, by reweighting with $\Theta$, as 
\be
\langle \Theta(z_{\eta},\overline{z}_{\eta}) O(z_{\eta})\rangle_{\eta}/\langle \Theta(z_{\eta},\overline{z}_{\eta}) \rangle_{\eta}.
\label{eq:CL_reweighting_formula}
\ee
Even when the random noise or equivalently $\hbar$ correction is included, 
so long as $P$ is well localized around each saddle point, this prescription seems to work nicely. 
Note that this replacement does not break DS equations (\ref{eq:holomorphic_DS_eq}) so far as the semiclassical analysis is valid. 

Now our question is ``What $d_{\sigma}$, or $c_{\sigma}$, is adopted in the complex Langevin method?"   
If $\re\, \omega_{\sigma}\le 0$ means $\langle \mathcal{K}_{\sigma},\mathbb{R}\rangle=0$, 
the following $d_{\sigma}$ is consistent with $c_{\sigma}\ge 0$:
\be
d_{\sigma}
=\sqrt{{\omega_{\sigma}\over |\omega_{\sigma}|}}\mathrm{e}^{\im\, \iim S_{\sigma}/\hbar}\langle \mathcal{K}_{\sigma},\mathbb{R}\rangle. 
\label{eq:conjecture_CL_Langevin_coefficient}
\ee
If this is true, the complex Langevin method gives an extension of the so-called phase quenched approximation to include complex saddle points: 
\be
\langle O(z)\rangle_{\eta}\simeq {1\over Z}\sum_{\sigma} \langle \mathcal{K}_{\sigma},\mathbb{R}\rangle \sqrt{2\pi \hbar \over |\omega_{\sigma}|}\mathrm{e}^{-\re\,S_{\sigma}/\hbar} O(z_{\sigma}). 
\ee
We adopt it as a working hypothesis in the following sections, although this is not the unique solution for consistency. 
Using this hypothesis, the phase function $\Theta$ is given by 
\be
\Theta(z,\overline{z}):= {1\over Z}\sum_{\sigma}\sqrt{|\omega_{\sigma}|\over \omega_{\sigma}}\mathrm{e}^{-\im\, \iim S_{\sigma}/\hbar} \chi_{\sigma}(z,\overline{z}), 
\ee
and the reweighting formula (\ref{eq:CL_reweighting_formula}) is available for practical use\footnote{ 
A similar improvement of complex Langevin method by reweighting with saddle-point phases has been discussed in Ref.~\cite{Fujii}.}.

\section{Numerical simulation}
We test our proposal by applying it to two models with and without a singular drift term.
We numerically solved \eqref{eq:Langevin} with the fictitious time step $\varepsilon=5.0\times10^{-6}$ and  $5.0\times10^{-7}$ for the models with and without the singular drift term, respectively.
We adopted a higher order algorithm~\cite{Aarts:2011zn}.  
Errors were estimated by using the jackknife method, and each quantity is computed by using $5.0\times10^5$ configurations.
Below we set $\hbar=1$.

\subsection{One-site fermion model}
First, as a nontrivial example with a singular drift term, we analyze a one-site fermion model. 
This is the simplest model to suffer from the sign problem same as that in lattice QCD simulations~\cite{Tanizaki:2015rda,Fujii:2015bua, Fujii:2015vha, Alexandru:2015xva}.
To demonstrate that the modified complex Langevin method can simulate the Silver Blaze like feature~\cite{Cohen:2003kd} in the one-site model 
is a good landmark to show its applicability to the sign problem in many-body systems.

After introducing a Hubbard--Stratonovich field $\vphi$, we explicitly integrate out the original fermionic fields. 
The partition function reads~\cite{Tanizaki:2015rda}
\be
	Z=\int {\diff \varphi_{\mathrm{bg}}}\; \mathrm{e}^{-S(\varphi_{\mathrm{bg}})} ,
\label{Eq:ZeroModeIntegral_Hubbard}
\ee
with the action,
\begin{equation}
\begin{split}
S(\varphi_{\mathrm{bg}})=\frac{\beta}{2U}\varphi_{\mathrm{bg}}^2 - 2\ln \left(1+\mathrm{e}^{\beta\left(\im \varphi_{\mathrm{bg}}+\mu+U/2\right)}\right),
\end{split}
\label{Eq:Classical_Action_OneSiteHubbard}
\end{equation}
where $\varphi_{\mathrm{bg}}=\int_0^\beta\diff\tau\varphi(\tau)/\beta$ is the zero Matsubara mode of $\varphi$.
 $U(>0)$, $\mu$, and $\beta=1/T$ are the on-site repulsive interaction, chemical potential, and inverse temperature, respectively. 
We dropped nonzero Matsubara modes of $\varphi$, since they do not couple to $\mu$~\cite{Tanizaki:2015rda}.
The auxiliary field $\varphi_{\mathrm{bg}}$ is related to the fermion number density $n$ by 
\be
n=-\langle\partial S/\partial (\beta\mu)\rangle=\iim\left[\langle \varphi_{\mathrm{bg}}\rangle/U\right] ,
\ee 
where we used the equation of motion to obtain the last expression.
The integral~\eqref{Eq:ZeroModeIntegral_Hubbard} is analytically calculable, but instead, we shall apply the complex Langevin method. 
Due to the logarithmic term, the action has infinitely many saddle points, which appear in the period of $2\pi T$. 
Since the Lefschetz-thimble method is valid even with these logarithmic singularities~\cite{Tanizaki:2014tua, Kanazawa:2014qma}, 
all the discussions in previous sections are available in order to conclude the failure of the complex Langevin method. 

In the large $\beta U$ limit, 
the saddle points $\vphi_m$ are given as~\cite{Tanizaki:2015rda} 
\be
\vphi_{m}=\im\left(\mu+{U\over 2}\right)+T\left(2\pi m +\im\ln {{3\over2}U-\mu\over {1\over 2}U+\mu}\right)+O(T^2)
\label{Eq:ApproximateSaddlePoints}
\ee 
with $m\in\mathbb{Z}$. 
The classical action at $\vphi_m$ reads~\cite{Tanizaki:2015rda}
\begin{eqnarray}
&S_0\simeq-{\beta U\over 2}\left(\frac{\mu}{U}+{1\over 2}\right)^2,
\label{Eq:ApproximateClassicalAction_OneThimble}\\
&\mathrm{Re}\,\left[S_m-S_0\right]\simeq {2\pi^2\over \beta U}m^2, 
\label{Eq:ApproximateClassicalActionRe}\\
&\mathrm{Im}\,S_m\simeq 2\pi m\left({\mu\over U}+\frac{1}{2}\right).
\label{Eq:ApproximateClassicalActionIm} 
\end{eqnarray}
In Eqs.~\eqref{Eq:ApproximateClassicalActionRe} and \eqref{Eq:ApproximateClassicalActionIm}, we have calculated only the $m$-dependent leading terms in the large $\beta U$ expansion. 
In this model, $\beta^{-1}$ plays a role of $\hbar$ but the classical action (\ref{Eq:Classical_Action_OneSiteHubbard}) depends on it in a nontrivial way. 
Therefore, $\mathrm{Re}[S_m]$ becomes a function of $\beta$ and $m$, which make difficult to judge the dominance of saddle points. 
According to Eq.~\eqref{Eq:ApproximateClassicalActionRe}, saddle points with $m^2\lesssim \beta U$ would give dominant contributions in the large $\beta U$ limit. 
Thus, the zero temperature limit, which corresponds to the classical limit in Sec.~\ref{Sec:FailuerOfComplexLangevin},  
is not described by the unique saddle point
and the condition $d_{\sigma}=\langle \mathcal{K}_{\sigma},\mathbb{R}\rangle$ is not trivially recovered.
According to Eq.~\eqref{Eq:ApproximateClassicalActionIm}, these different saddle points have different complex phases, and thus the complex Langevin simulation may fail except for special cases.

\begin{figure}[t]
\centering
\includegraphics[scale=0.85]{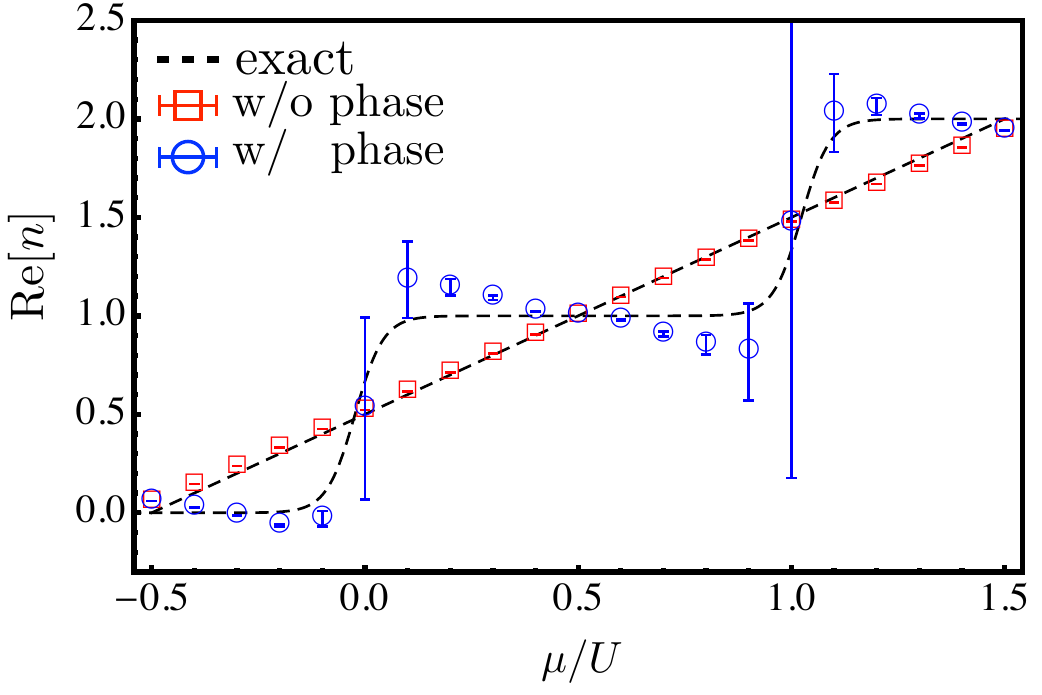}
\caption{Fermion number $n$ as a function of chemical potential $\mu$ at $\beta U=30$ with and without reweighting.}
\label{fig:hubbard_density}
\end{figure}

We show the fermion number $n$ as a function of the chemical potential $\mu$ in Fig.~\ref{fig:hubbard_density}.
The satndard complex Langevin method predicts the wrong linear $\mu$-dependence. This wrong behavior is also obtained from the mean field or the one-thimble approximation~\cite{Tanizaki:2015rda}. 
To find a reweighting factor, we use approximate expressions on the saddle points in the leading order of the large $\beta U$ expansion in Eqs.~\eqref{Eq:ApproximateClassicalAction_OneThimble}-\eqref{Eq:ApproximateClassicalActionIm}~\cite{Tanizaki:2015rda}.
The saddle points are in between singular points of the logarithm $\vphi^s_m=\im\left(\mu+{U\over 2}\right)+T\left(2\pi m +1/2\right)$, namely, $\re\,\vphi^s_m<\re\,\vphi_m<\re\,\vphi^s_{m+1}$.
The distribution generated by solving the complex Langevin equation is localized arournd $\vphi_m$ and decays by a power law as it is getting close to $\vphi^s_m$ along the real part direction.
For the imaginary part direction the distribution exponentially decays. 
Then we put $\chi_m(\vphi_{\rm bg},\overline{\vphi}_{\rm bg})=\theta(\re\, \vphi^s_m <\re\, \vphi_{\rm bg}<\re\, \vphi^s_{m+1} )$.
The residual sign coming from $\omega_\sigma$ turns out to be negligible for $\beta U=30$. 
Now $\Theta$ is given explicitly as 
\be
\Theta=\sum_{m\in\mathbb{Z}} \mathrm{e}^{-2\pi\im(\mu/U+1/2)m} \,\theta(2m-1 <\beta\,\re\, \vphi_{\rm bg}/\pi<2m+1 ),
\ee
where $\theta(x)$ is the step function.
We also show the fermion number after reweighting in Fig.~\ref{fig:hubbard_density}. 
The result becomes much better, but we may still need an improvement of $\Theta$ for exact agreement. 
The number density seems to linearly decrease in each plateaux as chemical potential increases.
This behavior is incorrect from the view point of the thermodynamics stability since the compressibility must be non-negative. 
There might exist the physics not included in our weighting prescription.

\begin{figure}[t]
\centering
\includegraphics[scale=.95]{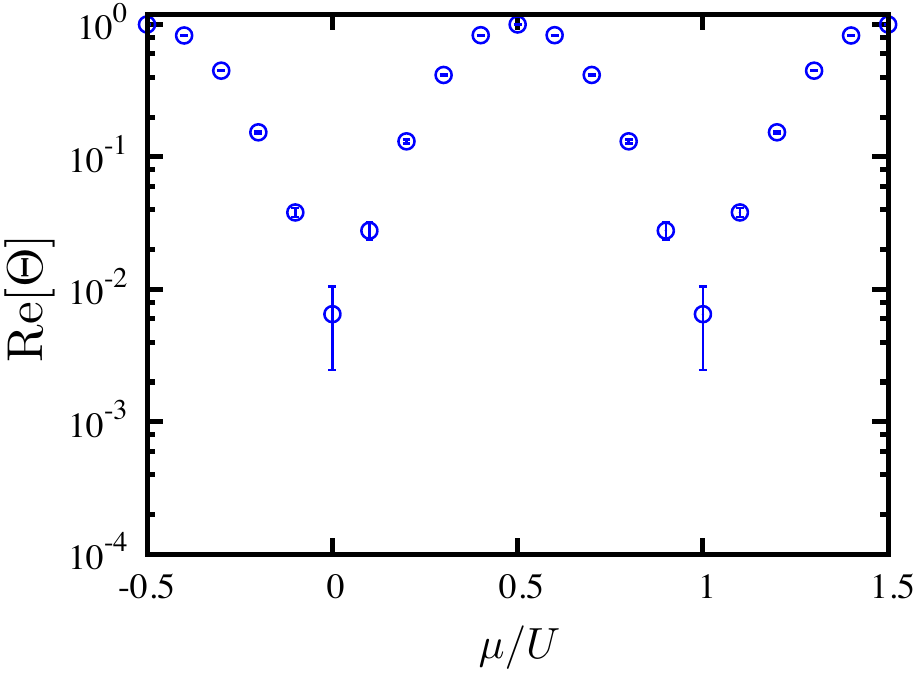}
\caption{Average phase function $\langle \Theta(z_{\eta},\overline{z}_{\eta})\rangle_{\eta}$ as a function of chemical potential $\mu$ at $\beta U=30$.}
\label{fig:weight}
\end{figure}

We show the average phase factor $\langle \Theta\rangle_{\eta}$ as a function of the chemical potential $\mu$ in Fig.~\ref{fig:weight}.
It becomes small near jumping points of $n$ at $\mu/U=0$ and $1$, and is getting close to one near the half filling $\mu/U=1/2$. 
If we apply the conventional reweighting by the Monte Carlo method to the original integral~\eqref{Eq:ZeroModeIntegral_Hubbard}, however, the severe sign problem appears for every $\mu/U>-1/2$~\cite{Tanizaki:2015rda}. 
The cancellation of the phase function in the modified complex Langevin method is milder than that in the reweighting by the Monte Carlo method.
The same cancellation may happen near phase transition points in many-body systems.
If $\langle \Theta\rangle_{\eta}$ becomes exponentially small as the system size increases, 
it is also true that the sign problem is still obstinate in the complex Langevin method. 

\subsection{Double-well potential model}
Next, we consider a model without a singular drift term, whose action is given by 
\be \label{Eq:Polynomial_Model}
S(x)=x^4/4-x^2/2-\im \alpha x,
\ee
with $\alpha>0$.  
This action has three saddle points on the complex plane.
Only two of them have positive $\re\, \omega_\sigma $, and contribute to the semiclassical analysis. 
These two saddle points ($z_1$ and $z_2$) are, respectively, located on the first and second quadrant planes ($\re\, z_1>0$ and $\re\,z_2<0$). 
They have different complex phases, except when $\alpha=0$. 
The complex Langevin simulation may fail at finite $\alpha$ from our semiclassical analysis given in Sec.~\ref{Sec:FailuerOfComplexLangevin}.

\begin{figure}[t]
\centering
\includegraphics[scale=.95]{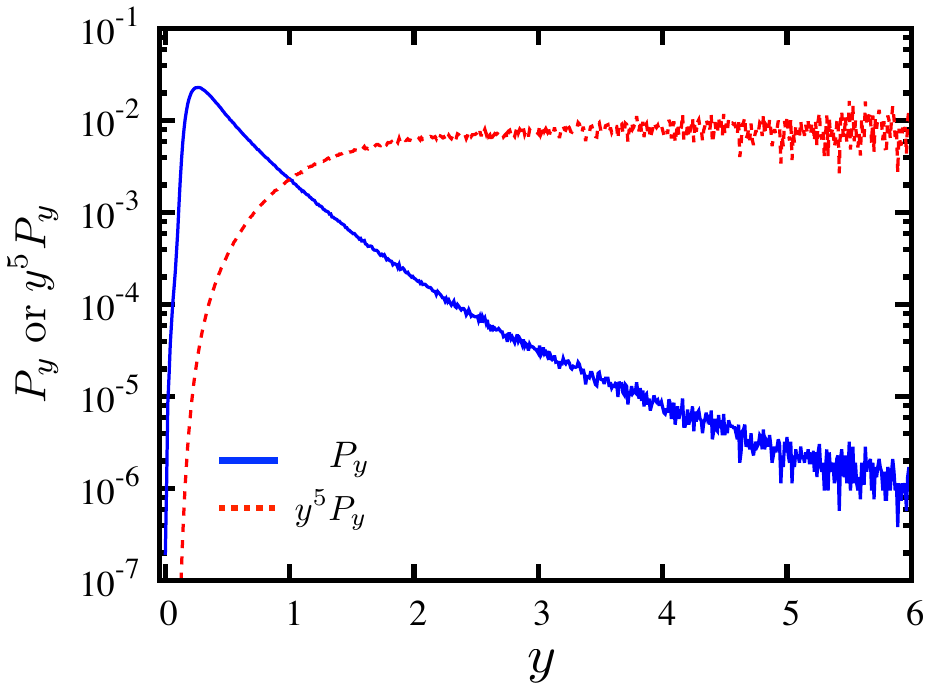}
\caption{Partially integrated distribution $P_y$ and its moment $y^5P_y$ at $\alpha=1.4$.
}
\label{fig:Py}
\end{figure}

The distribution of this model seems to have the power law behavior.
We show the partially integrated distribution
\be \label{Eq:distribution_y}
P_y(y)=\int\diff x \,P(x,y)/\int\diff x\diff y \,P(x,y) ,
\ee
and its fifth moment $y^5P_y$ in Fig.~\ref{fig:Py}.
The distribution may behave as $P_y\sim y^{-5}$ at $y\to\infty$.
The power law implies that the expectation value of a higher power of $z$, e.g., $z^n$ ($n\geq$4) diverges\footnote{Recently, it is mathematically shown that the power law is always true for any polynomial model if we use the complex noise instead of the real one~\cite{Herzog2014Noise1}. For the real noise, it seems to depend on a model whether the distribution shows the power law.}. 
Thus the complex Langevin simulation apparently breaks down, as expected.
Remark here that since $z^n$ ($n\geq$1) does not satisfy the DS equation~\eqref{eq:holomorphic_DS_eq} if the power law exponent is true,
our argument based on the DS equation in Sec.~\ref{Sec:FailuerOfComplexLangevin} is no longer available. 
Nevertheless the complex Langevin method actually breaks down, and our prescription works well for lower dimensional operators as is seen in the following.

\begin{figure}[t]
\centering
\includegraphics[scale=0.85]{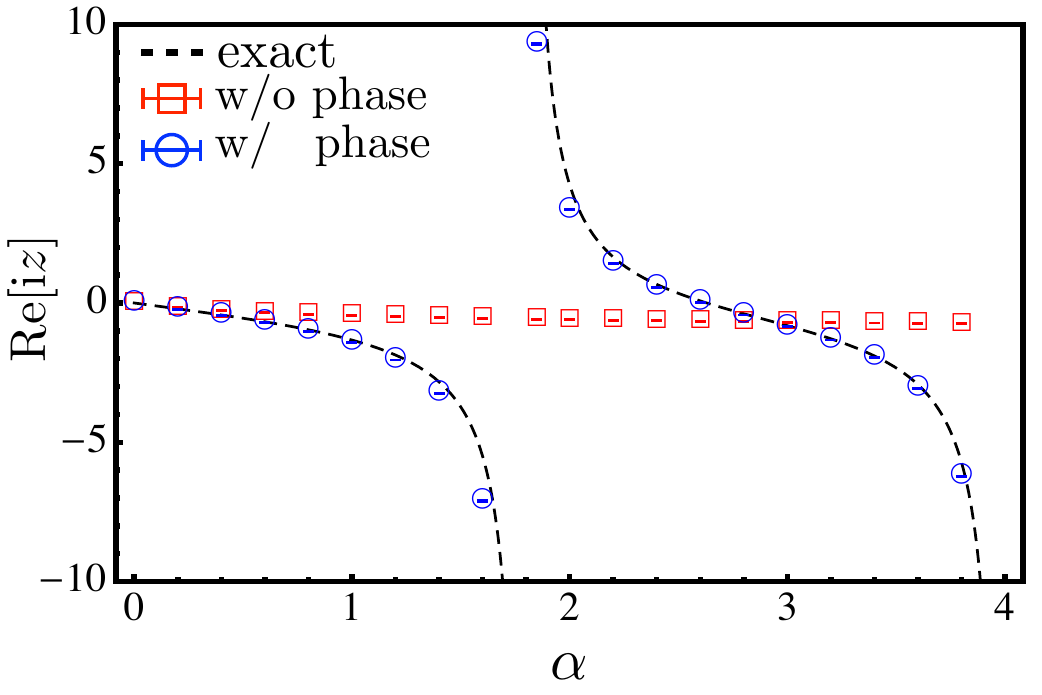}
\caption{Expectation value of $\im z$ as a function of $\alpha$, with and without reweighting. 
}
\label{fig:dwp_x}
\end{figure}
\begin{figure}[ht]
\centering
\includegraphics[scale=0.85]{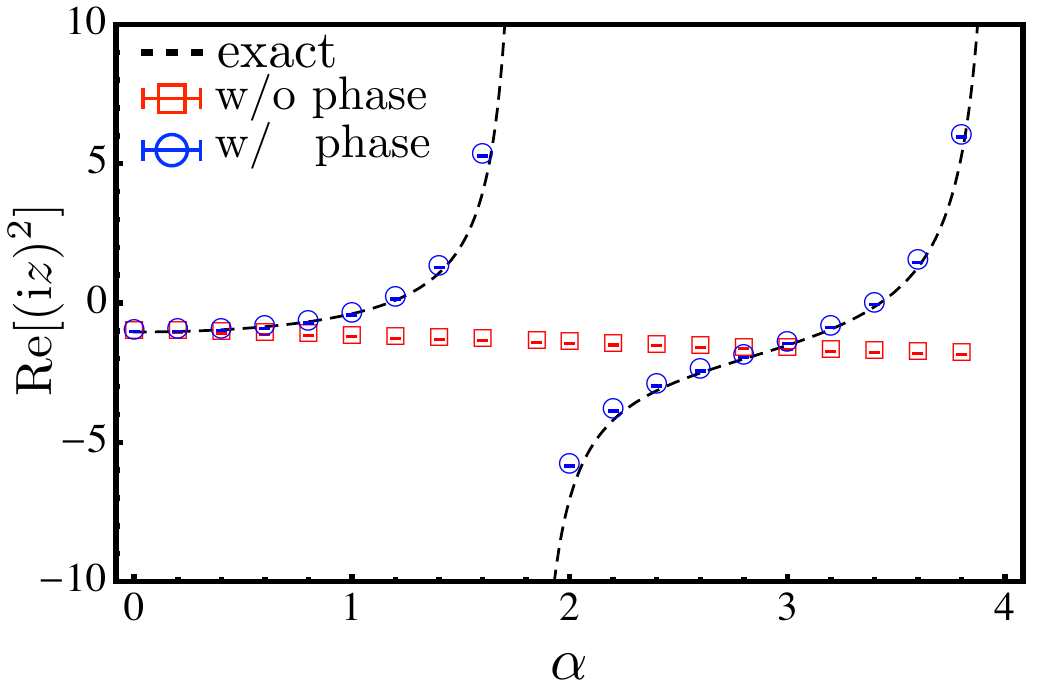}
\caption{Expectation value of $(\im z)^2$ as a function of $\alpha$, with and without reweighting.}
\label{fig:dwp_x2}
\end{figure}

We show the expectation values of $\im z$ and $(\im z)^2$ as a function of $\alpha$ in Figs.~\ref{fig:dwp_x} and~\ref{fig:dwp_x2}. 
The complex Langevin simulation converges to a wrong result (red squares.) 
Based on our prescription, we put $\chi_1(z,\overline{z})=\theta(\re\, z)$ and $\chi_2(z,\overline{z})=\theta(-\re\, z)$.
The result of the reweighting is also shown in Figs.~\ref{fig:dwp_x} and~\ref{fig:dwp_x2} with blue circles.
The reweighting works perfectly, 
and we resolve the wrong convergence problem. 
This is also true for $(\im z)^3$.
For a diverging higher power of $z$, $(\im z)^n$ ($n\geq$4), our prescription does not work, 
and the expectation values suffer from the large fluctuations before and after reweighting.

\section{Concluding remarks}

We have analytically shown within the semiclassical approximation
that complex Langevin method gives wrong results, when there are several dominant saddle points with different complex phases. 
Since the interference of these complex phases is an essential ingredient to understand the Silver Blaze phenomenon \cite{Tanizaki:2015rda}, the usual complex Langevin method might not be reliable in order to tackle the cold and dense nuclear matters. Moreover, this interference is also of great importance in order to study the dynamical phenomena, such as a particle production, using the real-time path integral~\cite{Dumlu:2010ua, Dumlu:2011rr, Dumlu:2011cc}.
For more general situation where the semiclassical analysis breaks down, 
we need further study to show the failure of the complex Langevin method. 

The next problem is to modify the distribution so as to reproduce the expectation values in the original theory.
We proposed a reweighting prescription by introducing a working hypothesis, which is consistent with the semiclassical analysis. 
This must be justified or revised in future study. Also the correct treatment of subdominant saddle points must be clarified.
Our prescription is numerically confirmed for two models with and without singular drift terms. 
In particular, we succeeded to simulate the nonanalytic behavior of the one-site fermion model at low temperatures.  

If our prescription were proven or revised, the modified complex Langevin method could provide a way to perform numerical simulations on multiple Lefschetz thimbles. 
However, it requires us to get complete knowledge on complex saddle points to assign correct phase function. 
Furthermore, our prescription causes a large cancellation of relative phases among saddle points, 
although it is somewhat milder than that of the  conventional reweighting by the Monte Carlo method.
This implies the sign problem possibly occurs in the modified complex Langevin method.
To find more efficient prescription must be an important future study. 

\section*{Acknowledgments}
T.H. thanks A.~Yamamoto for stimulating discussions.
Y.T. was supported by Grants-in-Aid for the fellowship of Japan Society for the Promotion of Science (JSPS) (No.25-6615) and is supported by Special Postdoctoral Researchers Program of RIKEN. 
Y.H. is partially supported by JSPS KAKENHI Grants Numbers 15H03652.
This work was partially supported by the RIKEN interdisciplinary Theoretical Science (iTHES) project, and by the Program for Leading Graduate Schools of Ministry of Education, Culture, Sports, Science, and Technology (MEXT), Japan.

\providecommand{\href}[2]{#2}\begingroup\raggedright\endgroup

\end{document}